\documentclass[english,12pt]{article}
\usepackage[latin1]{inputenc}
\usepackage{babel}
\usepackage{ngerman}
\usepackage{amsmath,amssymb, mathrsfs, dsfont, amsthm, color}
\usepackage[margin=1in]{geometry}
\usepackage{tikz}

\usepackage{amsxtra}
\usepackage{stmaryrd}

\renewenvironment{thebibliography}[1]{%
\begin{oldthebibliography}{#1}%
\setlength{\baselineskip}{.9em}
\linespread{.9}
\small
\setlength{\parskip}{0ex}%
\setlength{\itemsep}{.1em}%
}%
{%
\end{oldthebibliography}%
}

\usepackage{amsfonts}
\usepackage{amsthm}
\usepackage{amsmath}
\usepackage{amssymb}
\usepackage{bbm}
\usepackage{dsfont}
\newtheorem{thm}{Theorem}[section]
\newtheorem{defi}[thm]{Definition}
\newtheorem{prop}[thm]{Proposition}

\newtheorem{lemma}[thm]{Lemma}

\theoremstyle{definition}
\newtheorem{remark}[thm]{Remark}

\newcommand{\bt}{\begin{thm}}
\newcommand{\et}{\end{thm}}
\newcommand{\br}{\begin{remark}}
\newcommand{\er}{\end{remark}}
\newcommand{\bl}{\begin{lemma}}
\newcommand{\el}{\end{lemma}}
\newcommand{\bp}{\begin{proof}}
\newcommand{\ep}{\end{proof}}
\newcommand{\bal}{\begin{align*}}
\newcommand{\eal}{\end{align*}}
\newcommand{\bi}{\begin{itemize}}
\newcommand{\be}{\begin{equation}}
\newcommand{\ee}{\end{equation}}
\newcommand{\bea}{\begin{eqnarray}}
\newcommand{\eea}{\end{eqnarray}}
\newcommand{\ba}{\begin{align*}}
\newcommand{\ea}{\end{align*}}
\newcommand{\ei}{\end{itemize}}

\DeclareMathOperator{\esssup}{ess\ sup}
\DeclareMathOperator{\essinf}{ess\ inf}

\newcommand{\R}{\mathbb{R}}

\newcommand{\N}{\mathbb{N}}

\newcommand{\F}{\mathcal{F}}
\newcommand{\cF}{\mathcal{F}}
\newcommand{\cU}{\mathcal{U}}

\newcommand{\cA}{\mathcal{A}}
\newcommand{\Om}{\Omega}
\newcommand{\om}{\omega}

\newcommand{\tvp}{\widetilde\varphi}
\newcommand{\hvp}{\widehat\varphi}
\newcommand{\hpsi}{\widehat\psi}

\newcommand{\vp}{\varphi}

\newcommand{\ve}{\varepsilon}

\newcommand{\dom}{\mathrm{dom}}

\newcommand{\ssint}{\stackrel{\mbox{\tiny$\scriptscriptstyle\bullet$}}{}}

\newcommand{\cD}{\mathcal{D}}

\newcommand{\tS}{\widetilde S}
\newcommand{\hS}{\widehat S}

\newcommand{\hY}{\widehat Y}

\newcommand{\sint}{\stackrel{\mbox{\tiny$\bullet$}}{}}

\newcommand{\cC}{\mathcal{C}}
\newcommand{\cY}{\mathcal{Y}}
\newcommand{\cB}{\mathcal{B}}

\numberwithin{equation}{section}
\newcommand{\Inf}{\inf\limits}
\newcommand{\Lim}{\lim\limits}

\begin{document}
\title{Transaction Costs, Shadow Prices, and Duality\\ in Discrete Time}
\author{Christoph Czichowsky\footnote{London School of Economics and Political Science,
Department of Mathematics,
Columbia House,
Houghton Street,
London WC2 2AE,
UK, {\tt c.czichowsky@lse.ac.uk}. Financial support by the Swiss National Science Foundation (SNF)
under grant PBEZP2\_137313 is gratefully acknowledged.} 
\hspace{20pt}Johannes Muhle-Karbe\footnote{Department Mathematik, ETH Z\"urich, R\"amistrasse 101, CH-8092 Z\"urich, and Swiss Finance Institute, {\tt johannes.muhle-karbe@math.ethz.ch}. Partially supported by the National Centre of Competence in Research Financial Valuation and Risk Management (NCCR FINRISK), Project D1 (Mathematical Methods in Financial Risk Management), of the Swiss National Science Foundation (SNF).}
\\
Walter Schachermayer\footnote{Fakult\"at f\"ur Mathematik, Universit\"at Wien, Oskar-Morgenstern-Platz 1, A-1090 Wien, {\tt walter.schachermayer@univie.ac.at}. Partially supported by the Austrian Science Fund (FWF) under grant P25815, the European Research Council (ERC) under grant FA506041 and by the Vienna Science and Technology Fund (WWTF) under grant MA09-003.}
}
\date{\today}
\maketitle

\begin{abstract}
\noindent

For portfolio choice problems with proportional transaction costs, we discuss whether or not there exists a \emph{shadow price}, i.e., a least favorable frictionless market extension leading to the same optimal strategy and utility.

By means of an explicit counter-example, we show that shadow prices may fail to exist even in seemingly perfectly benign situations, i.e., for a log-investor trading in an arbitrage-free market with bounded prices and arbitrarily small transaction costs.

We also clarify the connection between shadow prices and duality theory. Whereas dual minimizers need not lead to shadow prices in the above ``global'' sense, we show that they always correspond to a ``local'' version.

\end{abstract}
\vspace{-0.2cm}\newline
\vspace{-0.2cm}\newline

\newpage

\section{Introduction}
A fundamental question in the theory of portfolio choice with proportional transaction costs is whether or not there exists a \emph{shadow price}, i.e., a least favorable frictionless market extension that leads to the same optimal strategy and utility. If the answer is affirmative, this implies that the behavior of a given economic agent can be explained by passing to a suitable frictionless shadow market. Put differently, no qualitatively new effects arise due to the market frictions. If, on the other hand, a shadow price fails to exist then the impact of transaction costs cannot be reproduced in any frictionless market.

Answers to this existence question have proved to be rather elusive beyond finite probability spaces, where shadow prices always exist \cite{KMK11}. In an It\^o process setting, Cvitani\'c and Karatzas~\cite{CK96} proved that a shadow price exists and corresponds to the solution of a suitable dual problem, \emph{if} the latter is attained in a set of martingales. Yet, existence of a dual minimizer could later only be guaranteed in a larger set of supermartingales (cf.\ \cite{CW01}, also compare \cite{deelstra.al.01,CO11}), for which the interpretation as a least favorable market extension is not clear. Loewenstein~\cite{L00} showed that shadow prices exist in a continuous filtration if short positions are ruled out, and this condition is in fact sufficient for the existence of shadow prices for general filtrations (cf.\ Benedetti et al.~\cite{BCKMK11}). However, a counter-example in the last study shows that shadow prices may fail to exist without further assumptions. In parallel, a different counter-example has been put forward by Rokhlin \cite{R11}. He shows that shadow prices generally only exist for a suitable relaxation of the optimization problem at hand, where the original utility functional is replaced by its lower semicontinuous envelope.  

The above counter-examples crucially revolve around either non-standard utility functionals involving Banach limits \cite{R11}, or unbounded price jumps, unbounded (relative and absolute) bid-ask spreads, and a
 candidate shadow price that admits arbitrage \cite{BCKMK11}. It remained unclear which of these properties are indeed necessary for the non-existence of shadow prices. Conversely, neither Loewenstein~\cite{L00} nor Benedetti et al.~\cite{BCKMK11} explain why the presence of shortselling constraints leads to positive results. 

The present study sheds light on both of these questions. We present a counter-example showing that shadow prices can fail to exist even in seemingly perfectly benign settings, i.e., for a log-investor trading in a market with bounded prices and constant bid-ask spread of arbitrary size. We explain how this happens because transaction costs change the set of strategies with non-negative terminal positions. That is, shadow prices can fail to exist if investment choices with transaction costs are strictly more constrained than in any potential frictionless shadow market. This observation also explains why shortselling constraints remedy the problem: evidently, long-only portfolios always lead to solvent positions both with and without transaction costs. 

In addition, we also clarify the connection between shadow prices and solutions to the dual problem. Starting with the seminal work of Cvitani\'c and Karatzas \cite{CK96}, it has become ``folklore'' that -- morally speaking -- shadow prices should correspond to the minimizer of the corresponding dual problem. In the present study, we discuss this in a discrete-time setting to avoid technicalities and focus on the main ideas. The extension to continuous time leads to thorny conceptual and substantial technical difficulties, ranging from the ubiquitous issue of admissibility to delicate questions about path properties of the limits of martingales. It is therefore deferred to future research. As in \cite{CK96}, we find that if the minimizer among the dual supermartingales is in fact a martingale, then it corresponds to a shadow price. Conversely -- under minimal regularity assumptions -- the dual minimizers lead to the only potential shadow prices. In view of the counter-examples in \cite{BCKMK11,R11} as well as in the present study, the solution to the dual problem does not, in general, lead to a shadow price in the above ``global'' sense. However, it always lies in the superdifferential of the conditional value process, and hence can be interpreted as a kind of ``local'' version as follows. Suppose that, at any fixed trading date, the investor is allowed to carry out a single trade at the local frictionless shadow price derived from the dual minimizer. Then, even though this provides potentially more favorable terms of trade, the investor will not deviate from her optimal position in the financial market with frictions. Across several periods, however, this property breaks down due to different solvency constraints.  

The remainder of the article is organized as follows. In Section 2,  we describe our discrete-time setting, the utility maximization problem under proportional transaction costs, and the notion of a shadow price. Section 3 contains a discussion of our counter-example on an intuitive level; the rigorous mathematical derivations are deferred to Appendix A for better readability. Section 4 studies the dual minimization problem and its connections to the (non-)existence of shadow prices; the corresponding proofs are collected in Appendix B.

\section{Utility maximization with transaction costs}

\subsection{Preliminaries}
We consider a discrete-time financial market with one riskless and one risky asset.\footnote{Several risky assets can be treated along the same lines, but we restrict ourselves to a single one to ease notation.} The riskless asset can be traded without frictions and its price is assumed to be normalized to one. Trading the risky asset incurs proportional transaction costs $\lambda \in (0,1)$. This means that one has to pay a higher ask price $S_t$ when buying risky shares but only receives a lower bid price $(1-\lambda)S_t$ when selling them. Here, \mbox{$S=(S_t)_{t=0}^T$} denotes a strictly positive, adapted process defined on a discrete-time filtered probability space $\big(\Om,\F,(\F_t)_{t=0}^T,P\big)$ with fixed finite time horizon $T\in\N$. We assume that $\cF_0$ is trivial and write $\Delta X_t:=X_t-X_{t-1}$, as well as $\varphi \sint X_t=\sum_{s=1}^t \varphi_s \Delta X_s$ for the stochastic integral of $\varphi$ with respect to $X$. 

\emph{Trading strategies} are modeled by $\R^2$-valued, predictable processes $\vp=(\vp^0_t,\vp^1_t)_{t=0}^{T+1}$, where $\vp^0_{t+1}$ and $\vp^1_{t+1}$ describe the holdings in the riskless and the risky asset, respectively, after rebalancing the portfolio at time $t$. 
A strategy $\vp=(\vp^0_t,\vp^1_t)_{t=0}^{T+1}$ is called \emph{self-financing}, denoted by $\vp\in\cA$, if purchases and sales of the risky asset are accounted for in the riskless position:
\begin{equation}\label{eq:sf}
\Delta\varphi^0_{t+1}\leq-(\Delta\varphi^1_{t+1})^+ S_t +(\Delta\varphi^1_{t+1})^-(1-\lambda)S_t, \quad 0 \leq t \leq T.
\end{equation}
We restrict our attention to self-financing strategies starting from a strictly positive initial cash endowment and liquidating the risky position at the terminal time $T$, i.e., satisfying $(\varphi^0_0,\varphi^1_0)=(x,0)$ as well as $\varphi^0_{T+1}\geq0$ and $\varphi^1_{T+1}=0$. This set is denoted by $\cA(x)$.

\begin{remark}
Without transaction costs, the above notions reduce to their well-known frictionless counterparts. Indeed, if bid and ask prices coincide for $\lambda=0$, purchases and sales do not need to be tracked separately. Instead, one simply works with the number of risky shares $\varphi^1_t$. The corresponding self-financed wealth is then given by the stochastic integral $x+\varphi^1 \sint S_t$, and the associated riskless position evolves as $\varphi^0_t=x+ \varphi^1 \sint S_t - \varphi^1_t S_t$. See, e.g., \cite[Chapter 5.1]{FS04} for more details. In the present study, our main focus lies on the interplay between frictional and frictionless markets, cf.\ Section~\ref{sec:sp}. Hence, we restrict ourselves to the above setting, rather than working with a general cone-valued model as in \cite{CS06,kabanov.safarian.09} because this allows us to formulate all results in direct analogy to the frictionless case. 
\end{remark}

In the above discrete-time setting, we consider an investor whose preferences are modeled by a utility function $U:(0,\infty)\to\R$ in the usual sense.\footnote{That is, a strictly increasing and strictly concave function that is continuously differentiable and satisfies the Inada conditions $U'(0):=\Lim_{x\searrow 0}U'(x)=\infty$ and $U'(\infty):=\Lim_{x\nearrow \infty}U'(x)=0$.} She trades to maximize the expected utility from her terminal cash position:
\be
E[U(g)]\to\max!\label{P1}
\ee
Here, $g$ runs through the set $\cC(x)=\{\vp^0_{T+1}\in L^0_+(P)~|~\text{$\exists(\vp^0,\vp^1)\in\cA(x)$}\}$ of non-negative cash positions that can be self-financed with a strategy $(\varphi^0,\varphi^1)$ starting from the initial endowment $(x,0)$.

\subsection{Shadow prices}\label{sec:sp}

Consider \emph{any} fictitious risky asset that can be traded \emph{without frictions} ($\lambda=0$) at a price $\tS=(\tS_t)_{t=0}^T$ taking values in the bid-ask spread $[(1-\lambda)S,S]$ of the original market with transaction costs. Then, since purchases and sales can be carried out at potentially more favorable prices, any attainable payoff in the market with bid-ask spread $[(1-\lambda)S,S]$ can be dominated by a payoff in the frictionless market with price process $\tS$. Hence, 
\begin{align}
u(x):=\sup_{\vp^0_{T+1}\in\cC(x)}E[U(\vp^0_{T+1})]&\leq\inf_{\tS\in[(1-\lambda)S,S]}\sup_{\vp^0_{T+1}\in\cC(x; \tS)}E[U(\vp^0_{T+1})]\nonumber\\
&=:\inf_{\tS\in[(1-\lambda)S,S]}u(x;\tS),\label{int:rel2}
\end{align}
where $\cC(x; \tS)$ denotes the set of non-negative payoffs that can be attained by self-financing strategies $(\vp^0,\vp^1) \in \cA(x;\tS)$ for the frictionless price process $\tS$. The natural question that arises here is whether one can find some particularly unfavorable frictionless shadow market with the same maximal expected utility as the original market with transaction costs. To achieve this, the corresponding frictionless optimizer has to purchase resp.\ sell stocks only when $\tilde{S}$ matches the ask resp.\ bid price, because trades at strictly more favorable prices would lead to  higher utility. Then, the frictionless optimizer in this ``shadow market'' is also available in the original market with transaction costs and hence optimal there as well. This motivates the following notion:

\begin{defi}\label{def:shadow}
An adapted process $\hS=(\hS_t)_{t=0}^T$ is called a \emph{shadow price} if it takes values in the bid-ask spread $[(1-\lambda)S,S]$ and there is a solution $(\varphi^0,\varphi^1)$ to the corresponding frictionless utility maximization problem
\be
E[U(\vp^0_{T+1})]=E[U(x+\vp^1\sint \hS_{T})]\to\max!,\qquad (\varphi^0,\varphi^1)\in\cA(x; \hS),\label{p1}
\ee
which trades only at bid-ask prices: 
$$\{\Delta\varphi^1_{t+1}>0\}\subseteq\{\hS_t=S_t\}  \quad \mbox{and} \quad \{\Delta\varphi^1_{t+1}<0\}\subseteq\{\hS_t=(1-\lambda)S_t\}, \quad \mbox{for }  t=0,\ldots,T.$$
\end{defi}

Note that a shadow price $\hS=(\hS_t)_{t=0}^T$ depends on the process $S$, the investor's utility function, and on her initial endowment.

\section{A counter-example}\label{sec:example}

In this section, we present an example showing that shadow prices can fail to exist even in an arbitrage-free two-period market with bounded prices and arbitrarily small transaction costs. The discussion is kept on an informal level for better readability; the rigorous proofs are deferred to Appendix A.

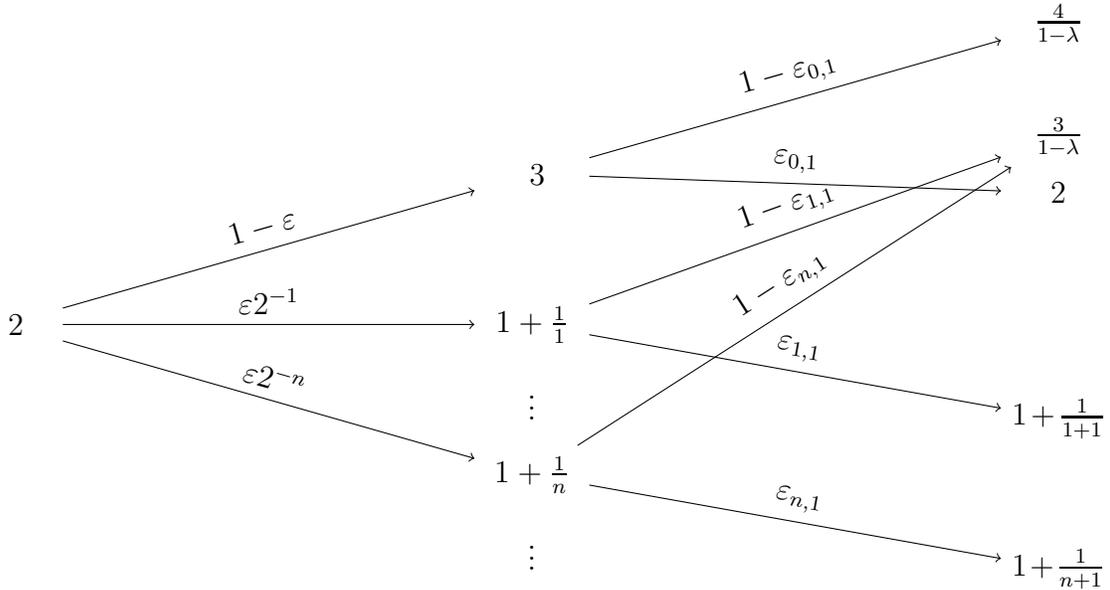
\begin{figure}[htbp]
 \tikzstyle{bag} = [text width=3em, text centered]
\tikzstyle{end} = []
\begin{center}
\begin{tikzpicture}[sloped]
 \node (a) at ( 0,0) [bag] {$\ \ 2$};
  \node (b00) at ( 7,-3) [bag] {$\vdots$};
 \node (b0) at ( 7,-2) [bag] {$1+\frac{1}{n}$};
 \node (b1) at ( 7,-1) [bag] {$\vdots$};
 \node (b2) at ( 7,0) [bag] {$1+\frac{1}{1}$};
 \node (b3) at ( 7,2) [bag] {\ 3};
\node (c1) at ( 14,-3.25) [bag] {$1+\frac{1}{n+1}$};
\node (c2) at ( 14,-1.25) [bag] {$1+\frac{1}{1+1}$};
\node (c3a) at ( 14,4) [bag] {$\frac{4}{1-\lambda}$};
\node (c3b) at ( 14,1.75) [bag] {$2$};
\node (c4) at ( 14,2.5) [bag] {$\frac{3}{1-\lambda}$};

\draw [->] (a) to node [above] {$\varepsilon 2^{-n}$} (b0);
\draw [->] (a) to node [above] {$\varepsilon 2^{-1}$} (b2);
\draw [->] (a) to node [above] {$1-\varepsilon$} (b3);
\draw [->] (b3) to node [above] {$1-\varepsilon_{0,1}$} (c3a);
\draw [->] (b3) to node [above] {$\varepsilon_{0,1}$} (c3b);
\draw [->] (b2) to node [above] {$1-\varepsilon_{1,1}$} (c4);
\draw [->] (b2) to node [above] {$\varepsilon_{1,1}$} (c2);
\draw [->] (b0) to node [above] {$1-\varepsilon_{n,1}$} (c4);
\draw [->] (b0) to node [above] {$\varepsilon_{n,1}$} (c1);
\end{tikzpicture}
\caption{Illustration of the ask price in the example. The nodes represent the values of the ask price at time $t=0,1,2$, connected by the corresponding transition probabilities. The parameters $\varepsilon,\varepsilon_{i,1}$ are chosen sufficiently small, cf.\ Appendix \ref{app:a} for more details.} \label{fig:shadow}
\end{center}
\end{figure}

The example is a variant of Example 5.1' in Kramkov and Schachermayer \cite{KS99}. Consider an investor with logarithmic utility $U(x)=\log(x)$ and an initial endowment of $x=1$ riskless and zero risky assets. The ask price $S_t$ of the risky asset evolves as illustrated in Figure \ref{fig:shadow} and the corresponding bid price is given by $(1-\lambda)S_t$ for a constant transaction cost $\lambda \in (0,1)$.

By choosing the constants $\varepsilon$ and $\varepsilon_{n,1}$, $n=0,1, \ldots$ in Figure \ref{fig:shadow} sufficiently small,\footnote{The formal definition is provided after Theorem \ref{thm:ex}.} downward moves of the risky asset become very unlikely, making it increasingly attractive for the investor to purchase as many shares of the risky asset as possible. In particular, the investor's positions in the risky asset will always be non-negative in this case.

On the other hand, the investor's holdings in the risky asset are limited by the constraint that she has to have a positive position after liquidating her portfolio at the terminal time $T$ to avoid bankruptcy. Since the bid price of the risky asset (at which the investor's non-negative risky position has to be liquidated) can go up or down in both periods, the liquidation value of the investor's portfolio has to be positive also at time $t=1$. Hence, the investor's acquisition $\Delta\varphi^1_1$ of risky assets in the first period has to satisfy the following solvency constraint:
$$1-\Delta\varphi^1_1 S_0+\Delta\varphi^1_1 (1-\lambda)S_1 \geq 0,$$
for all values of $S_1$. Since the latter can go down arbitrarily close to $1$, this implies that the investor's initial purchase $\Delta\varphi^1_1$ of risky assets is limited by $1/(1+\lambda)$. If the probabilities of downward movements are chosen small enough, this upper bound becomes binding, i.e., the investor buys the maximal number $\Delta\varphi^1_1=1/(1+\lambda)$ of risky shares at time $0$ that allows her to avoid bankruptcy. 

In the second period, the price of the risky asset can again go up or down. But now, the solvency constraint at time $t=2$ allows the investor to slightly increase her position in the risky asset in each of the states, which she will do if a suitable choice of the $\ve_{n,1}$ makes downward movements of the latter sufficiently unlikely. 

At the terminal time $t=2$, the investor just liquidates the positive number of risky assets she is holding.

In summary, the optimal strategy prescribes purchasing strictly positive amounts of the risky asset at times $t=0,1$, and liquidating the portfolio with a final sell trade at time $t=2$. Since a shadow price $\hS$ must by definition coincide with the trading prices whenever the optimal strategy transacts, the only \emph{candidate} for a shadow price $\hS$ is given by
$$\hS_0=S_0, \quad \hS_1=S_1, \quad \hS_2=(1-\lambda)S_2.$$
In contrast to the example of \cite{BCKMK11}, this frictionless market is arbitrage-free, and in fact can be shown to correspond to the minimizer of the dual problem (cf.\ Appendix A). Nevertheless, the corresponding optimal strategy for the frictionless market $\hS$ does \emph{not} coincide with the one in the original market with transaction costs. The reason is that in the frictionless shadow market the maximal number of shares that can be held in the first period is strictly bigger than in the original market with transaction costs. Indeed, any initial purchase $0\leq\Delta\varphi^1_1 \leq 1$ satisfies the frictionless solvency constraint $1 +\Delta\varphi^1_1 (\hS_1-\hS_0) \geq 0$. Hence, the investor can invest strictly more into the risky asset than in the original market with transaction costs (where the upper bound is $1/(1+\lambda)$), and this is indeed optimal if the probabilities of downward moves are chosen sufficiently small. Hence, the optimal strategy in the candidate shadow market $\hS$ strictly outperforms its frictional counterpart. Consequently, $\hS$ cannot be a shadow price and -- since it was the only candidate -- no shadow price exists in the example at hand. 

In fact, the above counter-example shows even more. Fix the above ask price for some $\lambda \in (0,1)$. Then, no shadow price exists for \emph{all} bid-ask spreads $\lambda' \in(0,\lambda)$. That is, a shadow price need not exist even for a fixed ask (or, equivalently, mid) price and arbitrarily small transaction costs. 

\section{Duality}\label{sec:duality}

We now discuss the connection between shadow prices and convex duality. To this end, we first formulate a suitable dual minimization problem (compare \cite{CK96,CW01,deelstra.al.01,CO11}) and show that the abstract versions of the results of Kramkov and Schachermayer \cite{KS99} allow to extend the key assertions of the frictionless theory also to the frictional case in a straightforward manner.

Fix a frictionless market $\tS=(\tS_t)_{t=0}^T$ and consider the primal utility maximization problem \eqref{int:rel2}. To formulate a corresponding dual minimization problem, consider the conjugate function $V$ of $U$, that is, the Legendre transform of $-U(-x)$ given by 
$$V(y):=\sup_{x>0}\{U(x)-xy\}, \quad \mbox{for $y>0$}.$$
In a frictionless market with price process $\tS$, \cite{KS99} show that a suitable domain of dual variables is given by
\begin{align*}
\cY(y;\tS)=\{&Y=(Y)_{t=0}^T \geq 0\ |\  Y_0=y \text{ and $Y(\varphi^0+\varphi^1 \tS)=\big(Y_t(\varphi^0_t+\varphi^1_t \tS_t)\big)_{t=0}^T$}\\
&\text{is a non-negative supermartingale for all $(\varphi^0,\varphi^1)\in\cA(1;\tS)$}\}.
\end{align*}
In the terminology of \cite{karatzas.kardaras.07} , this is the set of \emph{supermartingale deflators} that turn all admissible wealth process (in terms of the frictionless price process $\tS$) into supermartingales. Focusing on terminal values of the processes in $\cY(y;\tS)$, we define the set 
$$\cD(y;\tS)=\{h\in L^0_+(P)~|~\text{$\exists Y\in\cY(y;\tS)$ with $h\leq Y_T$}\}, \quad \mbox{for $y>0$},$$
of random variables dominated by a dual element. Then, the dual problem corresponding to the frictionless utility maximization problem is given by
\be
E[V(Y_T)]\to\min!,\qquad{Y\in\cY(y;\tS)},\label{d1}
\ee
or, equivalently,
$$E[V(h)]\to\min!,\qquad{h\in\cD(y;\tS)}.$$

With transaction costs, \emph{any} frictionless price process $\tS$ with values in the bid-ask spread can potentially be used to valuate the risky position in the investor's portfolio. At first glance, it is therefore appealing to use as dual variables the set of all such ``consistent price processes'' and, for each one, the set of associated supermartingale deflators. Unfortunately, this set is in general not large enough to contain the dual minimizer. Therefore, one has to work with the following slightly different notions:
\begin{align*}
\cB(y)=\big\{(Y^0,Y^1) \geq 0\ \big|\ &Y^0_0=y, \tfrac{Y^1}{Y^0} \in[(1-\lambda)S,S]\text{ and }Y^0(\vp^0+  \vp^1\tfrac{Y^1}{Y^0})=Y^0\vp^0+Y^1\vp^1  \\
&\text{is a non-negative supermartingale for all $(\vp^0,\vp^1)\in\cA(1)$}\big\},
\end{align*}
and, accordingly, 
$$\cD(y)=\{h\in L^0_+(P)~|~\text{$\exists (Y^0,Y^1)\in\cB(y)$ with $h\leq Y^0_T$}\}  \quad \mbox{for $y>0$}.$$
In the above definition of $\cB(y)$, the investor's risky position $\varphi^1$ is valued at the \emph{frictionless} price process $Y^1/Y^0$. However, multiplication with the supermartingale $Y^0$ only needs to turn those frictionless wealth processes $\varphi^0+\varphi^1 Y^1/Y^0$ into supermartingales that are generated by a strategy $(\varphi^0,\varphi^1)$ which is admissible in the original market with transaction costs. Since this is a (potentially strict) subset of the set of admissible strategies in the frictionless market with price process $Y^1/Y^0$, the process $Y^0$ need \emph{not} be an admissible dual variable for the frictionless market with price process $Y^1/Y^0$. This subtle distinction disappears if the dual variable $(Y^0,Y^1)$ is a martingale (cf.\ the proof of Lemma 4.5), but plays a crucial role in the examples where shadow prices do not exist (compare Appendix A). This is why we use the above definitions of $\cB(y)$ and $\cD(y)$.

\begin{remark}
Let us compare these definitions with the notion of \emph{consistent price systems}~\cite{S04} for transaction costs $\lambda$ (henceforth $\text{CPS}^{\lambda}$) ubiquitous in the theory of no-arbitrage with transaction costs (cf., e.g., \cite{jouini.kallal.95,kabanov.stricker.01,kabanov.al.02,S04} as well as the monograph of Kabanov and Safarian \cite{kabanov.safarian.09} and the references therein). The set $\mathcal{Z}$ of {\it $\lambda$-consistent price systems} consists of those martingales $(Z^0,Z^1) \geq 0$ such that the ratio $Z^1/Z^0$ evolves within the bid-ask spread $[(1-\lambda)S,S]$. Since its first component is the density process of an equivalent martingale measure for the frictionless risky asset $\tS:=Z^1/Z^0$, existence of a $\text{CPS}^{\lambda}$ rules out arbitrage also in the original market $S$ with transaction costs $\lambda$, since the terms of trade are never more favorable in the latter. The Fundamental Theorem of Asset Pricing with Transaction Costs then states that the converse also holds true true (see, e.g., \cite{kabanov.stricker.01,S04}); in this sense, $\text{CPS}^{\lambda}$ play a role analogous to martingale measures in the frictionless case. One readily verifies that the set $\mathcal{Z}$ of $\lambda$-consistent price systems is contained in $\cB(y)$ if the initial value of the first component is rescaled by $y$. Hence, the set $\cB(y)$ of dual variables is an extension of the set of $\text{CPS}^\lambda$, just as martingale measures are contained in the set of supermartingale deflators used in the frictionless case. Indeed, the set of dual variables is precisely the Fatou-closure of the smaller sets in both cases.\footnote{Compare the discussion in \cite[p.\ 47]{S04b} for the frictionless case.} More specifically, the proof of Lemma \ref{lpolar} below shows that the set $\mathcal{B}(y)$ of supermartingale densities in the definition of the dual variables $\cD(y)$ can be replaced by almost sure limits of $\text{CPS}^\lambda$: let
\begin{align*}
\widetilde{\cB}(y)=\Big\{Y=(Y^0_t,Y^1_t)_{t=0}^T~\Big|~&\text{$\exists(Z^{0,n},Z^{1,n})\in\mathcal{Z}$ such that}\\
&\text{$Y^0_t=\lim_{n\to\infty} yZ^{0,n}_t$ and $Y^1_t=\lim_{n\to\infty} yZ^{1,n}_t$ for each $t=0,\ldots,T$}\Big\}.
\end{align*}
Then, $\cD(y)=\{h\in L^0_+(P)~|~\text{$\exists (Y^0,Y^1)\in\widetilde{\cB}(y)$ with $h\leq Y^0_T$}\}$ for $y>0$.
\end{remark}

Using the dual variables proposed above, we have the following duality results in direct analogy to the frictionless case \cite[Theorems 2.1 and 2.2]{KS99}. For better readability, the proof is deferred to Appendix B. 

\bt\label{mainthm}
Suppose that $S$ satisfies $(CPS^{\lambda'})$ for some $\lambda'\in[0,\lambda)$,\footnote{This is equivalent to the ``robust no-arbitrage property'' of Schachermayer \cite{S04}, i.e., absence of arbitrage even for slightly smaller trading costs.} the asymptotic elasticity of $U$ is strictly less than one, i.e., $AE(U):=\limsup\limits_{x\to\infty}\frac{xU'(x)}{U(x)}<1$, and the maximal expected utility is finite, $u(x):=\sup_{g\in\cC(x)}E[U(g)]<\infty$, for some $x\in(0,\infty)$. Then:
\bi
\item[{\bf 1)}] The primal value function $u$ and the dual value function
$$v(y):=\inf_{h\in\cD(y)}E[V(h)]$$
are conjugate, i.e.,
\begin{eqnarray*}u(x)=\inf_{y>0}\{v(y)+xy\},\qquad v(y)=\sup_{x>0}\{u(x)-xy\},
\end{eqnarray*}
and continuously differentiable on $(0,\infty)$. The functions $u$ and $-v$ are strictly concave, strictly increasing, and satisfy the Inada conditions
$$\text{$\Lim_{x\to0}u'(x)=\infty,\qquad\Lim_{y\to\infty}v'(y)=0,\qquad\Lim_{x\to\infty}u'(x)=0,\qquad\Lim_{y\to0}v'(y)=-\infty$}.$$
\item[{\bf 2)}] For all $x,y>0$, the solutions $\widehat g (x)\in\cC(x)$ and $\widehat h(y)\in\cD(y)$ to the primal problem
\begin{equation*}
\textstyle
E\left[U(g)\right]\to\max!, \qquad{g\in\cC(x)},
\end{equation*}
and the dual problem
\begin{equation}
\textstyle
E\left[V(h)\right]\to\min!\label{D1}, \qquad{h\in\cD(y)},
\end{equation}
exist, are unique, and there are
$\big(\hvp^0(x),\hvp^1(x)\big)\in\cA(x)$ and $\big(\widehat{Y}^0(y),\widehat{Y}^1(y)\big)\in\cB(y)$ such that
\be
\text{$\hvp^0_{T+1}(x)=\widehat g(x)\qquad$ and $\qquad\widehat{Y}^0_T(y)=\widehat h(y)$.}\label{martcond}
\ee
\item[{\bf 3)}] For all $x>0$, let $\widehat y (x)=u'(x)>0$ which is the unique solution to
$$
v(y)+xy\to\min!,\qquad y>0.
$$
Then, $\widehat g (x)$ and $\widehat h \big(\widehat y(x)\big)$ are given by $(U')^{-1}\big(\widehat h \big(\widehat y(x)\big)\big)$ and $U'\big(\widehat g (x)\big)$, respectively, and we have that $E\big[\widehat g(x)\widehat h\big(\widehat y (x)\big)\big]=x\widehat y(x)$. In particular, the process
$$\widehat{Y}^0\big(\widehat y(x)\big)\hvp^0(x)+ \widehat{Y}^1\big(\widehat y(x)\big)\hvp^1(x)=\Big(\widehat{Y}^0_t\big(\widehat y(x)\big)\hvp^0_t(x)+ \widehat{Y}^1_t\big(\widehat y(x)\big)\hvp_t^1(x)\Big)_{t=0}^T$$
is a martingale for all $\big(\hvp^0(x),\hvp^1(x)\big)\in\cA(x)$ and $\big(\widehat{Y}^0\big(\widehat y(x)\big),\widehat{Y}^1\big(\widehat y(x)\big)\big)\in\cB\big(\widehat y (x)\big)$ satisfying \eqref{martcond} with $y=\widehat y (x)$.
\item[{\bf 4)}] 
Moreover, 
$$\widehat{Y}^0\big(\widehat y(x)\big)\hvp^0(x)+ \widehat{Y}^1\big(\widehat y(x)\big)\hvp^1(x)=\widehat{Y}^0\big(\widehat y(x)\big)\left(x+ \hvp^1(x)\sint \tfrac{\widehat{Y}^1}{\widehat{Y}^0}\right),$$
which implies that $\{\Delta\hvp^1_{t+1}>0\}\subseteq\left\{\frac{\widehat{Y}^1_t}{\widehat{Y}^0_t}=S_t\right\}$ and $\{\Delta\hvp^1_{t+1}<0\}\subseteq\left\{\frac{\widehat{Y}^1_t}{\widehat{Y}^0_t}=(1-\lambda)S_t\right\}$ for $t=0,\ldots,T$.
\item[{\bf 5)}] Finally, we have $v(y)=\Inf_{(Z^0,Z^1)\in\mathcal{Z}}E[V(yZ^0_T)]$.
\ei
\et

\begin{remark}
Note that while the terminal values $\hvp^0_{T+1}(x)$ and $\widehat{Y}^0_T(y)$ are unique, this is not necessarily true for the processes $\big(\hvp^0(x),\hvp^1(x)\big)$ and $\big(\widehat{Y}^0(y),\widehat{Y}^1(y)\big)$. For the optimal strategies, this is in analogy to the frictionless case, where uniqueness also only holds for the optimal terminal payoff. For the dual minimizer, uniqueness is guaranteed for the first component $\hat{Y}^0$ by strict concavity of the dual objective function like in the frictionless case. However, the second component need not be unique. E.g., if the mid-price $S$ is a martingale, any martingale evolving in the bid-spread (e.g., any suitable multiple of the mid price) will lead to a dual minimizer after multiplying with $\hat{Y}^0$.
\end{remark}

\emph{Under the assumptions of Theorem \ref{mainthm}}, we have the following two results clarifying the connection between dual minimizers and shadow prices. The proofs are again postponed to Appendix B. The first result shows that the following insight of Cvitani\'c and Karatzas~\cite{CK96} is also true in the present context: if there is no ``loss of mass'' in the dual problem with transaction costs, then its minimizer corresponds to a shadow price:

\begin{prop}\label{lem:martingale}
If there is a minimizer $(\widehat{Y}^0,\widehat{Y}^1)\in\cB\big(\widehat{y}(x)\big)$ of the dual problem \eqref{d1} which is a martingale,\footnote{Note that, since we are working in finite discrete time, every non-negative local martingale is actually a true martingale.} then $\hS:=\widehat{Y}^1/\widehat{Y}^0$ is a shadow price.
\end{prop}

Conversely, the following result shows that \emph{if} a shadow price exists, it is necessarily derived from a dual minimizer.

\begin{prop}\label{lem:connection}
If a shadow price $\hS$ exists, it is given by $\hS=\widehat{Y}^1/\widehat{Y}^0$ for a minimizer $(\widehat{Y}^0,\widehat{Y}^1)\in\cB\big(\widehat{y}(x)\big)$ of the dual problem \eqref{d1}.
\end{prop}

In view of the counter-example presented in Section \ref{sec:example}, the dual minimizer from Theorem \ref{mainthm} does not necessarily give rise to a frictionless shadow market in the \emph{global} sense of Definition \ref{def:shadow}. \emph{Locally} for each period $(t,t+1)$, however, it can always be interpreted in such a manner. To make this idea precise, we consider the two-dimensional optimization problem corresponding to \eqref{P1}, given by
\be
E[\cU(\vp^0_{T+1},\vp^1_{T+1})]\to\max!,\qquad{(\vp^0,\vp^1)\in\cA(x)},\label{P2}
\ee
with the two-dimensional objective function
$$\mathcal{U}(x,y):=\begin{cases} U(x)~&:\text{ $y\geq 0$ and $x>0$},\\
-\infty~&:\text{ else}.
\end{cases}$$
For every $t=0,\ldots, T$ and every $(\psi^0,\psi^1)\in L^0(\Om,\cF_t,P;\R^2)$, we define the \emph{$\cF_t$-conditional value function}:\footnote{Here, we use a generalized conditional expectation: setting $+\infty-\infty=-\infty$, we define 
\begin{equation*}
\begin{split}
E\big[\cU\big(\psi^0+(\tvp^0_{T+1}-\tvp^0_{t}),\psi^1+(\tvp^1_{T+1}-\tvp^1_{t})\big)\big|\cF_t\big]:=&E\big[\big(\cU\big(\psi^0+(\tvp^0_{T+1}-\tvp^0_{t}),\psi^1+(\tvp^1_{T+1}-\tvp^1_{t})\big)\big)^+\big|\cF_t\big]\\ &-E\big[\big(\cU\big(\psi^0+(\tvp^0_{T+1}-\tvp^0_{t}),\psi^1+(\tvp^1_{T+1}-\tvp^1_{t})\big)\big)^-\big|\cF_t\big].
\end{split}
\end{equation*}}

$$\cU_t(\psi^0,\psi^1):=\underset{(\tvp^0,\tvp^1)\in\cA}{\esssup}E\big[\cU\big(\psi^0+(\tvp^0_{T+1}-\tvp^0_{t}),\psi^1+(\tvp^1_{T+1}-\tvp^1_{t})\big)\big|\cF_t\big].$$
All dual minimizers then lie in the $\cF_t$-conditional superdifferential along the optimal strategy $(\hvp^0,\hvp^1)$ of this $\cF_t$-conditional concave function in the following sense:

\begin{prop}\label{lem:super}
For any dual minimizer $(\widehat{Y}^0,\widehat{Y}^1)\in\cB\big(\widehat{y}(x)\big)$, we have that
\be
\cU_t(\psi^0,\psi^1)\leq \cU_t(\hvp^0_{t},\hvp^1_{t})+(\psi^0-\hvp^0_{t})\hY^0_{t}+(\psi^1-\hvp^1_{t})\hY^1_{t}\label{eq:supdiff}
\ee
a.s.~for all $(\psi^0,\psi^1)\in L^0(\Om,\cF_t,P;\R^2)$ and all $t=0,\ldots, T$.
\end{prop}
This result can be interpreted as follows. Suppose that up to time $t-1$, the investor has followed the strategy $\widehat\varphi$. At time $t$ she is allowed to make an additional trade in terms of the \emph{frictionless} price $\widehat{Y}^1_t/\widehat{Y}^0_t$ corresponding to the dual minimizer. If she trades a number $\nu$ of shares, her risky position then moves to $\psi^1_t:=\widehat\varphi^1_t +\nu$, whereas her cash balance changes to $\psi^0_t:=\widehat{\varphi}^0_t-\nu \widehat{Y}^1_t/\widehat{Y}^0_t$. Inequality \eqref{eq:supdiff} implies that none of these trades increases the investor's indirect utility: 
\begin{equation*}\label{eq:deviate}
\cU_t(\psi^0_{t},\psi^1_{t}) \leq \cU_t(\hvp^0_{t},\hvp^1_{t}).
\end{equation*}
Consequently, allowing the investor to trade in terms of the potentially more favorable frictionless price $\widehat{Y}_t^1/\widehat{Y}_t^0$ at a \emph{single} trading time $t$ does not cause her to deviate from the frictional optimal policy. Note, however, that in contrast to Definition \ref{def:shadow} this interpretation is only ``local'': the value functions on both sides of the above inequality still refer to the transaction cost problem in all other periods. In the counter-example from Section \ref{sec:example}, for example, increasing the risky investment in the first period beyond the value of the frictional optimizer would still lead to bankruptcy when liquidating at bid-ask prices later on, unlike when considering -- simulataneously in \emph{all} periods --  the  ``global'' frictionless market $\widehat{Y}^1/\widehat{Y}^0$ derived from the dual minimizer.

\appendix
\section{Rigorous analysis of the counter-example}\label{app:a}

This section contains a precise mathematical analysis of the counter-example discussed on an informal level in Section 3. To this end, we first compute the optimal policy in the original market with transaction costs by discrete-time dynamic programming, i.e., backward induction, and explicit calculations. This in turn determines a unique candidate shadow price, for which backward induction again leads to the optimal policy. The latter, however, does not coincide with the original one, and leads to higher utility. In summary, this shows:

\bt\label{thm:ex}
Fix $\lambda\in(0,1)$. There exists an arbitrage-free bounded process $S=(S_t)_{t=0}^2$ based on a countable probability space \mbox{$\Om=\{\om_{n,1}, \om_{n,2}\}_{n=0}^\infty$} with the following properties:
\begin{enumerate}
\item[{\bf 1)}] For the log-utility maximization problem with initial endowment $(\vp^0_{0},\vp^1_{0})=(1,0)$ and under transaction costs $\lambda \in (0,1)$, the solution $(\hvp^0_t,\hvp^1_t)_{t=0}^3$ to the primal problem
\be
E[U(\vp^0_3)]=E[\log(\vp^0_3)]\to\max!,\qquad{(\vp^0,\vp^1)\in\cA(1)},\label{logP1}
\ee
 and the solution $(\widehat Y^0_t,\widehat Y^1_t)_{t=0}^2$ to the dual problem
 \be
E[V(Y^0_2)]=E[-\log(Y^0_2)-1]\to\min!,\qquad{(Y^0,Y^1)\in\cB(1)},\label{logD1}
\ee
for $\widehat y(x)=1$ exist and are unique.
\item[{\bf 2)}] The unique candidate  $(\widehat S_t)_{t=0}^2:=(\widehat Y^1_t/\widehat Y^0_t)_{t=0}^2$ for a frictionless shadow price process $\hS$ is arbitrage-free, and takes values in the bid-ask spread $[(1-\lambda)S,S]$.
\item[{\bf 3)}] Despite 1) and 2), the frictionless log-utility optimization problem
 \be
E[\log(\varphi^0_3)]=E[\log(1+\varphi^1\sint \hS_2)]\to\max!,\quad(\varphi^0, \varphi^1) \in\cA(1; \hS),\label{p1log}
\ee
for $\hS$ yields a different solution and a higher value than its counterpart for $S$ under transaction costs~$\lambda$.
\end{enumerate}
\et

The remainder of Appendix A is devoted to proving Theorem \ref{thm:ex} by means of the explicit counter-example discussed on an informal level in Section \ref{sec:example}. To make the latter -- illustrated in Figure \ref{fig:shadow} -- precise, define the transition probabilities $P[\{\om_{0,1},\om_{0,2}\}]=(1-\ve)$, $P[\{\om_{0,1}\}]=(1-\ve)p_{0,1}$, $P[\{\om_{n,1}, \om_{n,2}\}]=\ve2^{-n}=:p_n$ and $P[\{\om_{n,1}\}]=p_n(1-\ve_{n,1})$, for \emph{sufficiently small} $\varepsilon,\varepsilon_{i,1}$. More specifically, we take $\ve\in(0,\frac{1}{3})$ and
\begin{align*}
p_{0,1}&:=(1-\ve_{0,1})=\frac{(1+2 \lambda ) (3+q_0+\lambda +q_0 \lambda )}{2 (1+\lambda ) (2+\lambda )},\\
p_{n,1}&:=(1-\ve_{n,1})=\frac{(1+n (2+n) \lambda ) \left((2 n-1) q_1 (1+\lambda )+n^2 (2+\lambda )\right)}{n^2 (1+n \lambda ) (1+2 \lambda +n (2+\lambda ))}
\end{align*}
with $q_0\in(0,\frac{1-\lambda }{1+3\lambda +2\lambda ^2})$ and $q_1\in(0,\frac{1-\lambda }{1+4 \lambda +3 \lambda ^2})$. To determine the optimal strategy in this example, apply discrete-time dynamic programming (cf., e.g., \cite{EK79} and the references therein) to the two-dimensional optimization problem corresponding to \eqref{logP1}:
\be
E[\cU(\vp^0_3,\vp^1_3)]\to\max!,\qquad{(\vp^0,\vp^1)\in\cA(1)},\notag\label{logP2}
\ee
with the two-dimensional objective function
$$\mathcal{U}(x,y):=\begin{cases} \log(x)~&:\text{ $y\geq 0$ and $x>0$},\\
-\infty~&:\text{ else}.
\end{cases}$$
The corresponding \emph{value function} is then given by
$$\cU_t(x,y)=\underset{(\tvp^0,\tvp^1)\in\cA}{\esssup}E\big[\cU\big(x+(\tvp^0_3-\tvp^0_{t}),y+(\tvp^1_3-\tvp^1_{t})\big)\big|\cF_t\big],$$
for $(x,y)\in\R^2$ and $t=0,1,2$. Here we define $\cU_t(x,y)=-\infty$, if there exists no $(\tvp^0,\tvp^1)\in\cA$ such that $\big(x+(\tvp^0_3-\tvp^0_{t}),y+(\tvp^1_3-\tvp^1_{t})\big)\in\dom(\cU)$. Since the bid and ask prices can go up or down in any period, this happens precisely if the \emph{liquidation value} $\ell_{t}(x,y):=x+y^+ (1-\lambda)S_{t}-y^{-} S_{t}$ of the position $(x,y)$ at time $t$ is nonpositive.

 By the definition of $\cU_t$ and the structure of $\cA$, we have the \emph{dynamic programming property}
\be
\cU_t(x,y)=\underset{\ell_t(-\Delta\tvp^0_{t+1},-\Delta\tvp^1_{t+1})\geq0}{\esssup}E[\cU_{t+1}\big(x+\Delta\tvp^0_{t+1},y+\Delta\tvp^1_{t+1})|\cF_t],\label{DPP}
\ee
which allows us to compute the solution $(\hvp^0,\hvp^1)$ to \eqref{logP1} by optimizing in \eqref{DPP} recursively.

\begin{remark}
$\ell_t(-\Delta\tvp^0_{t+1},-\Delta\tvp^1_{t+1})\geq0$ in \eqref{DPP} means that the trade at time $t$ needs to be self-financing. The second requirement of admissibility, namely a strictly positive liquidation wealth, is ensured automatically for the optimizer in \eqref{DPP} by definition of $\mathcal{U}$.
\end{remark}

\bl\label{lsollog}
The solution $(\hvp^0,\hvp^1)$ to problem \eqref{logP1} is given by $(\hvp^0_{0},\hvp^1_{0})=(1,0)$, 
$$\Delta\hvp^1_{1}=\frac{1}{1+\lambda}, \quad \Delta\hvp^1_{2}(\om_{0,i})=q_0, \quad \Delta\hvp^1_{2}(\om_{n,i})=\frac{q_1}{n}, \mbox{  for $i=1,2$ and $n\in\N$}, \quad \Delta\hvp^1_{3}=-\hvp^1_2,$$
where the cash position $\hvp^0$ is determined by the self-financing condition \eqref{eq:sf} (with equality). The superdifferential $\partial \cU(\hvp^0,\hvp^1)$ of the value function along the optimal strategy $(\hvp^0,\hvp^1)$ is given by
 \begin{align*}
\partial \cU_0(\hvp^0_0,\hvp^1_0)=\left\{\textstyle\frac{(1,S_0)}{\hvp^0_{0}+\hvp^1_{0}S_0}\right\}, \quad \partial \cU_1(\hvp^0_1,\hvp^1_1)=\left\{\textstyle\frac{(1,S_1)}{\hvp^0_{1}+\hvp^1_{1}S_1}\right\}, \quad \partial \cU_2(\hvp^0_2,\hvp^1_2)&=\left\{\textstyle\frac{(1,(1-\lambda)S_2)}{\hvp^0_{2}+\hvp^1_{2}(1-\lambda)S_2}\right\}.
\end{align*}
\el

\bp 
Since the investor's portfolio is liquidated at the terminal time $T=3$, 
\be
\cU_2(\vp^0_2,\vp^1_2)=\log\big(\vp^0_2+(\vp^1_2)^+(1-\lambda)S_2-(\vp_2^1)^-S_2\big)\label{supcU2}
\ee
for all $(\vp^0,\vp^1)\in\cA(1)$. At time $t=1$, we have
$$E[\mathcal{U}_2(\vp^0_1+\Delta\tvp^0_2,\vp^1_1+\Delta\tvp^1_2)|\F_1](\om_{n,i})=f_n(\Delta \tvp^1_2;\vp^0_1,\vp^1_1)$$
for $\Delta\tvp^1_2\geq0$ and $\vp^1_1+\Delta\tvp^1_2\geq0$, where
$$f_n(\Delta \tvp^1_2;\vp^0_1,\vp^1_1):=E\Big[\log\Big(\vp^0_1+\vp^1_1S_1+(\vp^1_1+\Delta\tvp^1_2)\big((1-\lambda)S_2-S_1\big)\Big)\Big|\F_1\Big](\om_{n,i})$$
for $i=1,2$ and $n\in\N_0$. Therefore the maximizer as well as the optimal value in \eqref{DPP} agree with their counterparts for
\be
f_n\big(\Delta \tvp^1_2(\om_{n,i});\vp^0_1,\vp^1_1\big)\to\max!,\qquad{\Delta \tvp^1_2(\om_{n,i})\in\R},\label{ap1}
\ee
for $i=1,2$ and $n\in\N_0$ as long as the maximizer of \eqref{DPP} falls into the domain where both functions coincide. (Note that local maxima are global maxima by concavity.) Since $f_n(\Delta \tvp^1_2(\om_{n,i});\vp^0_1,\vp^1_1)$ is differentiable \mbox{with respect to $\Delta\tvp^1_2(\om_{n,i})$,} the maximizer to \eqref{ap1} is determined by the first order condition \mbox{$f'_n\big(\Delta \tvp^1_2(\om_{n,i});\vp^0_1,\vp^1_1\big)=0$.} Solving the latter equation, we obtain that
\be
\Delta \hvp^1_2(\vp^0_1,\vp^1_1,0):=\big(\vp^0_1+\vp^1_1S_1(\om_{0,i})\big)\frac{1+\lambda}{\lambda+2}\left(\frac{1}{1+\lambda}+q_0\right)-\vp^1_1\label{solap0}
\ee
and
\be
\Delta \hvp^1_2(\vp^0_1,\vp^1_1,n):=\big(\vp^0_1+\vp^1_1S_1(\om_{n,i})\big)\frac{1+\lambda}{\lambda+\frac{1}{n}}\left(\frac{1}{1+\lambda}+\frac{q_1}{n}\right)-\vp^1_1,\quad\text{$n\in\N$},\label{solap1}
\ee
maximize \eqref{ap1} due to our choice of $p_{n,1}$.

Since
$$\big(\vp^0_1+\vp^1_1S_1(\om_{0,i})\big)\geq\big(\vp^0_1+\vp^1_1S_1(\om_{n,i})\big)\geq\frac{\lambda+\frac{1}{n}}{1+\lambda}$$
for all $i=1,2$, $n\in\N_0$, and $(\vp^0,\vp^1)\in\cA(1)$ with $\vp^1_1>0$ because $S_1(\om_{0,i})\geq S_1(\om_{n,i})$ and $\tvp^1_1\in\left(-\frac{1}{1+\lambda},\frac{1}{1+\lambda}\right]$ for all $(\vp^0,\vp^1)\in\cA(1)$, we obtain that $\Delta \hvp^1_2(\vp^0_1,\vp^1_1,n)\geq 0$ and $\vp^1_1+\Delta \hvp^1_2(\vp^0_1,\vp^1_1,n)\geq0$ for all $n\in\N_0$ and $(\vp^0,\vp^1)\in\cA(1)$ with $\vp^1_1>0$. Therefore $\Delta \hvp^1_2(\vp^0_1,\vp^1_1,n)\geq 0$ and $\vp^1_1+\Delta \hvp^1_2(\vp^0_1,\vp^1_1,n)\geq0$ for all $n\in\N_0$ for all $(\vp^0,\vp^1)\in\cA(1)$, as this is clearly also true for all $(\vp^0,\vp^1)\in\cA(1)$ with $\vp^1_1\leq0$.
Plugging $\Delta \hvp^1_2(\vp^0_1,\vp^1_1)$ into \eqref{DPP} gives
\be
\cU_1(\vp^0_{1},\vp^1_{1})=\log(\vp^0_{1}+\vp^1_{1}S_1)+C_1\label{cU1}
\ee
for all $(\vp^0,\vp^1)\in\cA(1)$, where 
$$
C_1 = 
\begin{cases}
E\left[\log \left(1+\frac{1+\lambda}{\lambda+2}\left(\frac{1}{1+\lambda}+q_0\right)\right)((1-\lambda)S_2-S_1)|\mathcal{F}_1\right](\omega_{0,\cdot}),\\
E\left[\log \left(1+\frac{1+\lambda}{\lambda+1/n}\left(\frac{1}{1+\lambda}+\frac{q_1}{n}\right)\right)((1-\lambda)S_2-S_1)|\mathcal{F}_1\right](\omega_{n,\cdot}).
\end{cases}
$$

At time $t=0$, we observe similarly as at time $t=1$ that
$$E[\mathcal{U}_1(\vp^0_0+\Delta\tvp^0_1,\vp^1_0+\Delta\tvp^1_1)]=h(\Delta \tvp^1_1;\vp^0_0,\vp^1_0)$$
for $\Delta\tvp^1_1\geq0$ and $\vp^1_0+\Delta\tvp^1_1\geq0$, where
$$h(\Delta\tvp^1_1;\vp^0_0,\vp^1_0):=E[\log(\vp^0_0+\vp^1_0S_0+(\vp^1_0+\Delta\tvp^1_1)(S_1-S_0))+C_1].$$
Then as before the maximizer as well as the optimal value in \eqref{DPP} agree with their counterparts for
\be
h(\Delta \tvp^1_1;\vp^0_0,\vp^1_0)\to\max!,\qquad{\Delta \tvp^1_1\in\left(-\frac{1}{1+\lambda},\frac{1}{1+\lambda}\right]},\label{ap0}
\ee
as long as the maximizer of \eqref{ap0} falls into the domain where both functions coincide. The function $h$ is differentiable with respect to $\Delta \tvp^1_1$ and its derivative is
$$h'(\Delta\tvp^1_1;\vp^0_0,\vp^1_0)=(1-\ve)\frac{1}{\vp^0_0+\vp^1_02+\Delta\tvp^1_1}+\ve\sum_{n=1}^\infty2^{-n}\frac{\frac{1}{n}-1}{\vp^0_0+\vp^1_02+\Delta\tvp^1_1(\frac{1}{n}-1)}.$$
For $\vp^0_0=1$, $\vp^1_0=0$ and $\Delta \tvp^1_1=\frac{1}{1+\lambda}$ we obtain that
\begin{align*}
h'\Big(\frac{1}{1+\lambda};1,0\Big)&=(1-\ve)\frac{2}{1+\frac{1}{1+\lambda}}+\ve\sum_{n=1}^\infty2^{-n}\frac{\frac{1}{n}-1}{1+\frac{1}{1+\lambda}(\frac{1}{n}-1)}\\
&\geq(1+\lambda)\left((1-\ve)\frac{1}{2}+\ve\sum_{n=1}^\infty2^{-n}(1-n)\right)\\
&=(1+\lambda)\left((1-\ve)\frac{1}{2}+\ve\left(1-\frac{\frac{1}{2}}{(\frac{1}{2}-1)^2}\right)\right)=(1+\lambda)\left(\frac{1}{2}-\ve\frac{3}{2}\right)>0
\end{align*}
for $\ve\in(0,\frac{1}{3})$, which implies that the concave function $h(\, \cdot\, ;1,0)$ attains its maximum over $\Delta \tvp^1_1\in(-\frac{1}{1+\lambda},\frac{1}{1+\lambda}]$ at $\Delta \tvp^1_1=\frac{1}{1+\lambda}$. Insertion  into \eqref{DPP} gives
\be
\cU_0(\vp^0_0,\vp^1_0)=\log(\vp^0_0+\vp^1_0S_0)+c_0\label{cU0}
\ee
with $c_0=E\big[\log\big(1+\frac{1}{1+\lambda}(S_1-S_0)\big)+C_1\big]$ for all $(\vp^0,\vp^1)\in\cA(1)$.

To verify the optimality of the strategy defined in the assertion, we observe that
$$E[\cU(\hvp^0_3,\hvp^1_3)]=\cU_0(1,0)=\sup_{(\vp^0,\vp^1)\in\cA(1)}E[\cU(\vp^0_3,\vp^1_3)]$$
by the above considerations, which already implies that $(\hvp^0,\hvp^1)$ is the solution to \eqref{logP2}.

As $\hvp^1_2>0$, \eqref{supcU2} yields that $\partial \cU_2(\hvp^0_2,\hvp^1_2)=\left\{\textstyle\frac{(1,(1-\lambda)S_2)}{\hvp^0_{2}+\hvp^1_{2}(1-\lambda)S_2}\right\}$. To compute $\partial \cU_1(\hvp^0_1,\hvp^1_1)$, we observe that \eqref{solap0} and \eqref{solap1} are continuous with respect to $(\vp^0,\vp^1)$ and that $\Delta \hvp^1_2(\hvp^0_1,\hvp^1_1,n)> 0$ and $\hvp^1_1+\Delta \hvp^1_2(\hvp^0_1,\hvp^1_1,n)>0$ for all $n\in\N_0$. Therefore \eqref{cU1} is also valid in some sufficiently small open neighborhood of $(\hvp^0_1,\hvp^1_1)$ and we obtain that $\partial \cU_1(\hvp^0_1,\hvp^1_1)=\left\{\textstyle\frac{(1,S_1)}{\hvp^0_{1}+\hvp^1_{1}S_1}\right\}$ by differentiating \eqref{cU1}. For $\vp=(\vp^0,\vp^1)\in\cA$ such that $(\vp^0_0,\vp^1_0)$ lies in some sufficiently small open neighborhood of $(\hvp^0_0,\hvp^1_0)=(1,0)$, we have that $\vp$ is solvent at time $t=1$ if $\Delta\vp^1_1\in\left(-\frac{\vp^0_0+\vp^1_0S_0}{1+\lambda},\frac{\vp^0_0+\vp^1_0S_0}{1+\lambda}\right]$, and therefore that \eqref{cU0} also holds in this case. Differentiating \eqref{cU0} then implies $\partial \cU_0(\hvp^0_0,\hvp^1_0)=\left\{\textstyle\frac{(1,S_0)}{\hvp^0_{0}+\hvp^1_{0}S_0}\right\}$, completing the proof.
\ep

\bl
The solution $(\hY^0,\hY^1)\in\cB(1)$ to the dual problem \eqref{logD1} is given by 
 \begin{align}
(\hY^0_0,\hY^1_0)=\frac{(1,S_0)}{\hvp^0_{0}+\hvp^1_{0}S_0}, \quad (\hY^0_1,\hY^1_1)=\frac{(1,S_1)}{\hvp^0_{1}+\hvp^1_{1}S_1}, \quad (\hY^0_2,\hY^1_2)&=\frac{(1,(1-\lambda)S_2)}{\hvp^0_{2}+\hvp^1_{2}(1-\lambda)S_2}.\label{solDP}
\end{align}
Both $\hY^0$ and $\hY^1$ are \emph{strict} supermartingales, i.e., fail to be true martingales. Nevertheless, $(\hS_t)_{t=0}^2:=(\hY^1_t/\hY^0_t)_{t=0}^2$ evolves in the bid-ask spread and the corresponding frictionless market is arbitrage-free.
\el

\bp
By Proposition \ref{lem:super} the solution to the dual problem is always valued in the $\cF_t$-conditional superdifferential of the $\cF_t$-conditional value function along the optimal strategy $(\hvp^0,\hvp^1)$. As this $\cF_t$-conditional superdifferential coincides in the present Markovian set-up with the superdifferential of the value function, which is by Lemma \ref{lsollog} single-valued, the process $(\hY^0,\hY^1)$ given in \eqref{solDP} is the unique solution to the dual problem \eqref{logD1}. 

To see that $\hY^0$ and $\hY^1$ are only supermartingales but no martingales, we note that \mbox{$E[\hY^0_1(S_1-S_0)|\F_0]=h'(\frac{1}{1+\lambda};1,0)\in\big(0,\frac{1+\lambda}{2}\big)$} and hence
$$E\bigg[\frac{\hY^0_1}{\hY^0_0}\bigg|\F_0\bigg]=1-\hvp^1_0E[\hY^0_1(S_1-S_0)|\F_0]<1,$$
which also implies that
$$E[\hY^1_1-\hY^1_0|\F_0]=E[\hY^0_1(S_1-S_0)+(\hY^0_1-\hY^0_0)S_0|\F_0]=E[\hY^0_1(S_1-S_0)|\F_0](1-\hvp^1_0S_0\hY^0_0)<0.$$

Since $(\hS_t)_{t=0}^2=(\hY^1_t/\hY^0_t)_{t=0}^2$ can go up and down with a positive probability in each period, $\hS$ is an arbitrage-free  price process. As $\hS$ takes values in the bid-ask spread $[(1-\lambda)S,S]$ by definition, this completes the proof.
\ep

The solution to the frictionless log-utility maximization problem for the price process $(\hS_t)_{t=0}^2:=(\hY^1_t/\hY^0_t)_{t=0}^2$  can again be computed recursively. The corresponding maximal expected utility turns out to be strictly higher than in the original market with transaction costs; hence $\hS$ cannot be a shadow price for the latter. Since it is the only candidate, this shows that a shadow price does not exist.

\bl\label{lsfp}
The solution $(\hpsi^0,\hpsi^1)$ to the frictionless log-utility maximization problem \eqref{p1log} with risky asset $\hS$ is given by $(\hpsi^0_0,\hpsi^1_0)=(1,0)$,
$$
\Delta\hpsi^1_1=1, \quad \Delta\hpsi^1_2(\om_{0,i})=2 \frac{1+\lambda}{2+\lambda}\left(\frac{1}{1+\lambda}+q_0\right)-1,\quad \Delta\hpsi^1_2(\om_{n,i})=\frac{1}{n}\frac{1+\lambda}{\lambda+\frac{1}{n}}\left(\frac{1}{1+\lambda}+\frac{q_1}{n}\right)-1,
$$
and $\Delta\hpsi^1_3=-\hpsi^1_2$ for $i=1,2$ and $n\in\N$, where the cash positions $\hpsi^0_t$, $t=1,2,3$ are determined by the self-financing condition \eqref{eq:sf} (with equality). The maximal expected utility in this frictionless market is strictly higher than in the original market with transaction costs,
$$\sup\limits_{(\varphi^0, \varphi^1) \in\cA(1; \hS)}E[\log(\varphi^0_3)]>\sup\limits_{(\vp^0,\vp^1)\in\cA(1)}E[\log(\varphi^0_3)].$$
Moreover, the frictionless dual minimizer $\widehat Y=1/(\widehat\psi^0+\widehat\psi^1 \hS)$ does not coincide with its frictional counterpart.
\el

\bp
The solution to the frictionless primal problem can be calculated recursively by dynamic programming as in Lemma \ref{lsollog}. At time $t=1$, one has to solve 
$$E\big[\log(1+\vp^1_1\Delta \hS_1+\vp^1_2\Delta \hS_2)\big|\F_1\big](\om_{n,i})=f_n(\vp^1_2;1+\vp^1_1\Delta \hS_1,0)\to\max!,\quad{\vp^1_2\in\R},$$
which, by \eqref{solap1}, has the solution $\hpsi^1_2(\om_{0,i})=(1+\vp^1_1\Delta \hS_1)\frac{1+\lambda}{\lambda+2}\left(\frac{1}{1+\lambda}+q_0\right)$ and $\hpsi^1_2(\om_{n,i})=(1+\vp^1_1\Delta \hS_1)\frac{1+\lambda}{\lambda+\frac{1}{n}}\left(\frac{1}{1+\lambda}+\frac{q_1}{n}\right)$ for $i=1,2$ and $n\in\N$. At time $t=0$, the frictionless $\hS$-agent can buy more stocks than the agent with transaction costs and therefore has the different admissibility constraint $\vp^1_1\in(-1,1]$. Hence, her optimal strategy is determined by solving the problem 
$$E[\log(1+\vp^1_{1}\Delta \hS_1)+C_1]=h(\vp^1_1;1,0)\to\max!,\quad{\vp^1_1\in(-1,1]}.$$
Similarly as in the proof of Lemma \ref{lsollog} we have $h'(1;1,0)=(\frac{1}{2}-\ve\frac{3}{2})>0$. Therefore the concave function $h(\,\cdot\,;1,0)$ attains its maximum over $(-1,1]$ at $\hpsi_1=1$ and 
$$\sup_{(\vp^0,\vp^1)\in\cA(\hS)}E[\log(1+\vp^1\sint \hS_2)]=h(\hpsi_1;1,0)>h(\Delta \hvp^1_0;1,0)=\sup_{(\vp^0,\vp^1)\in\cA(1)}E[\cU(\vp^0_3,\vp^1_3)].$$
Since $E[\log(1+\hpsi^1\sint \hS_2)]=\sup_{(\vp^0,\vp^1)\in\cA(\hS)}E[\log(1+\vp^1\sint \hS_2)]$, $(1+\hpsi^1\sint \hS_1)(\om_{0,i})=2$ and $(1+\hpsi^1\sint \hS_1)(\om_{n,i})=\frac{1}{n}$ for $i=1,2$ and $n\in\N$, the solution to \eqref{p1log} is given by $(\hpsi^0,\psi^1)$ as above. The frictionless dual minimizer is then given by $\hY=\frac{1}{1+\hpsi^1\ssint S}$, since $\hY_2=\frac{1}{1+\hpsi^1\ssint S_2}$ and $(1+\hpsi^1\sint S)\hY$ is a martingale by \cite[Theorem 2.2]{KS99}. This completes the proof.
\ep

A similar but even more tedious analysis of this example shows that a shadow price need not exist even if one fixes the ask price $S$ for some $\lambda \in (0,1)$, but considers all arbitrary small spreads $\lambda' \in(0,\lambda)$.

\section{Proofs for Section \ref{sec:duality}}

The proof of  parts 1)--3) of Theorem \ref{mainthm} follow from the abstract versions of the main results in \cite[Theorems 3.1 and 3.2]{KS99} once we have shown in the lemma below that the relations in \cite[Proposition 3.1]{KS99} hold true. We call a set $\mathcal{G}\subseteq L^0_+(P)$ \emph{solid}, if $0\leq f\leq g$ and $g\in\mathcal{G}$ imply that $f\in\mathcal{G}$, and use that $\cC(x)=x\cC(1)=:x\cC$ and \mbox{$\cD(y)=y\cD(1)=:y\cD$.} For part 5) of Theorem \ref{mainthm} we show in assertion 4) below that $D:=\{Z^0_T~|~(Z^0,Z^1)\in\mathcal{Z}\}$ is closed under countable convex combinations. Its proof then follows from \cite[Proposition 3.2]{KS99}.

\bl\label{lpolar}
Suppose that $S$ satisfies $(CPS^{\lambda'})$ for some $\lambda'\in(0,\lambda)$. Then:
\bi
\item[{\bf 1)}] $\cC$ and $\cD$ are convex, solid and closed in the topology of convergence in measure.
\item[{\bf 2)}] $g\in\cC$ iff $E[gh]\leq 1$, for all $h\in\cD$, and $h\in\cD$ iff $E[gh]\leq 1$, for all $g\in\cC$.
\item[{\bf 3)}] $\cC$ is a bounded subset of $L^0_+(P)$ and contains the constant function $1$.
\item[{\bf 4)}] $D:=\{Z^0_T~|~(Z^0,Z^1)\in\mathcal{Z}\}$ is closed under countable convex combinations.
\ei
\el
\bp
1) The sets $\cC$ and $\cD$ are convex and solid by definition. The assumption that $S$ satisfies $(CPS^{\lambda'})$ for some $\lambda'\in(0,\lambda)$ implies the existence of a strictly consistent price system (see Definition 1.5 in \cite{S04}). Therefore $\widehat{A}_T:=\{(\vp^0_T,\vp^1_T)~|~(\vp^0,\vp^1)\in\cA\}$ and hence \mbox{$\cC=\big((1,0)+\widehat{A}_T\big)\cap L^0_+(P)\times\{0\}$} are closed with respect to convergence in measure by Theorems 1.7 and 2.1 in \cite{S04}, where we identify $L^0(\Om,\cF,P;\R\times\{0\})$ with $L^0(\Om,\cF,P;\R)$. The closedness of $\cD$ follows by similar arguments as in Lemma 4.1 in \cite{KS99}. Indeed, let $(h^n)$ be a sequence in $\cD$ converging to some $h$ in measure. Then there exists a sequence $\big((Y^{0,n},Y^{1,n})\big)$ in $\cB(1)$ such that $Y^{0,n}_T\geq h^n$ for each $n\in\N$. Since $Y^{0,n}$ and $Y^{1,n}$ are non-negative supermartingales, there exist by (the arguments in the proof of) Lemma 5.2.1 in \cite{FK97} a sequence $(\widetilde{Y}^{n,0},\widetilde{Y}^{n,1})\in\mathrm{conv}\big((Y^{0,n},Y^{1,n}),(Y^{0,n+1},Y^{1,n+1}),\ldots\big)$ for $n\geq 1$ and supermartingales $\widetilde{Y}^0$ and $\widetilde{Y}^1$ such that $(\widetilde{Y}^{n,0}_t)$ is almost surely convergent to $\widetilde{Y}^0_t$ and $(\widetilde{Y}^{n,1}_t)$ to $\widetilde{Y}^1_t$  for each $t=0,\ldots,T$. This almost sure convergence is then sufficient to deduce that $(\widetilde{Y}^0,\widetilde{Y}^1)$ is $\R_+^2$-valued with $\widetilde{Y}^0_0=1$,  and that $\widetilde{Y}^0\vp^0+\widetilde{Y}^1\vp^1$ is a non-negative supermartingale for all $(\vp^0,\vp^1)\in\cA(1)$ and $\widetilde{Y}^0_T\geq h$, which implies that $(\widetilde{Y}^0,\widetilde{Y}^1)\in\cB(1)$ and hence that $h\in\cD$.

2) Since $D\subseteq\cD$, we obtain the first assertion by the superreplication theorem under transaction costs \cite[Theorem 4.1]{S04}. The second assertion then follows by the same arguments as the proof of Proposition 3.1 in \cite{KS99}.

3) The fact that $\cC$ contains the constant function $1$ follows by definition; the $L^0_+(P)$-boundedness is implied by the existence of a strictly positive element in $\cD$.

4) By martingale convergence, $D$ is closed in the norm of $L^1(P)$ and hence under countable convex combinations.
\ep
\bp[Proof of part 4) of Theorem \ref{mainthm}]
By the proof of part 1) of Lemma \ref{lpolar} above and part 5) of Theorem \ref{mainthm} there exists a sequence $\big((Z^{0,n},Z^{1,n})\big)$ in $\mathcal{Z}$ such that $\widehat{y}(x)Z^{0,n}_t$ and $\widehat{y}(x)Z^{1,n}_t$ converge almost surely to $\hY^0_t\big(\widehat{y}(x)\big)$ and $\hY^1_t\big(\widehat{y}(x)\big)$ for each $t=0,\ldots, T$. As $\frac{Z^{1,n}}{Z^{0,n}}$ is valued in the bid-ask spread $[(1-\lambda)S,S]$, any $(\vp^0,\vp^1)\in\cA(x)$ is also self-financing for $\frac{Z^{1,n}}{Z^{0,n}}$ without frictions ($\lambda=0$) and $Z^{0,n}\big(x+\vp^1\sint\frac{Z^{1,n}}{Z^{0,n}}\big)$ is a local martingale that is bounded from below and hence a supermartingale. Therefore we obtain that
\begin{align*}
E[Z^{0,n}_T\hvp^0_{T+1}(x)]&=\textstyle E\left[Z^{0,n}_T\left(\sum_{k=1}^{T+1}\left(\Delta \hvp^0_{k}(x)+\frac{Z^{1,n}_{k-1}}{Z^{0,n}_{k-1}}\Delta \hvp^1_{k}(x)\right)+x+\hvp^1(x)\sint \frac{Z^{1,n}}{Z^{0,n}}_{T}\right)\right]\\
&\leq\textstyle E\left[Z^{0,n}_T\left(\sum_{k=1}^{T+1}\left(\Delta \hvp^0_{k}(x)+\frac{Z^{1,n}_{k-1}}{Z^{0,n}_{k-1}}\Delta \hvp^1_{k}(x)\right)\right)\right]+x.
\end{align*}
Applying Fatou's Lemma twice the latter implies that
\begin{align*}
x\widehat{y}(x)&=E\big[\widehat{Y}^0_{T}\big(\widehat{y}(x)\big)\hvp_{T+1}^0(x)\big]\leq \liminf_{n\to\infty}E[\widehat y(x)Z^{0,n}_T\hvp^0_{T+1}(x)]\\
&=\liminf_{n\to\infty}\textstyle E\left[\widehat y(x)Z^{0,n}_T\left(\sum_{k=1}^{T+1}\left(\Delta \hvp^0_{k}(x)+\frac{Z^{1,n}_{k-1}}{Z^{0,n}_{k-1}}\Delta \hvp^1_{k}(x)\right)\right)\right]+x\widehat y(x)\\
&\leq\limsup_{n\to\infty}\textstyle E\left[\widehat y(x)Z^{0,n}_T\left(\sum_{k=1}^{T+1}\left(\Delta \hvp^0_{k}(x)+\frac{Z^{1,n}_{k-1}}{Z^{0,n}_{k-1}}\Delta \hvp^1_{k}(x)\right)\right)\right]+x\widehat y(x)\\
&\leq\textstyle E\left[\widehat{Y}^0_{T}\big(\widehat y(x)\big)\left(\sum_{k=1}^{T+1}\left(\Delta \hvp^0_{k}(x)+\frac{\widehat{Y}^1_{k-1}(\widehat y(x))}{\widehat{Y}^0_{k-1}(\widehat y(x))}\Delta \hvp^1_{k}(x)\right)\right)\right]+x\widehat y(x) 
\end{align*}
and therefore that $\Delta \hvp^0_{k}(x)+\frac{\widehat{Y}^1_{k-1}(\widehat y(x))}{\widehat{Y}^0_{k-1}(\widehat y(x))}\Delta \hvp^1_{k}(x)=0$ for $k=1,\ldots,T+1$, which yields the assertion.
\ep

\bp[Proof of Proposition \ref{lem:martingale}]
Suppose that $(\hY^0,\hY^1)\in\cB\big(\widehat y(x)\big)$ is a martingale. Then, the process $\hY^0\vp^0+ \hY^1\vp^1= \hY^0\big(x+\vp^1\sint\frac{\hY^1}{\hY^0}\big)$ is a non-negative local martingale and hence a supermartingale for all $(\vp^0,\vp^1)\in\cA\big(x;\frac{\hY^1}{\hY^0}\big)$,\footnote{Note that this argument fails in general if the dual minimizer is a strict supermartingale, unless negative risky positions are ruled out by short selling constraints as in \cite{L00,BCKMK11}} which implies that $\hY^0\in\cY\big(\widehat y(x);\frac{\hY^1}{\hY^0}\big).$ As $\hvp^0_{T+1}=(U')^{-1}(\hY^0_{T})$, we have $\hY^0_{T}=U'(\hvp^0_{T+1})$, and $\hY^0\hvp^0+ \hY^1\hvp^1= \hY^0\big(x+\hvp^1\sint\frac{\hY^1}{\hY^0}\big)$ is a martingale by Theorem \ref{mainthm}, we obtain by the duality for the frictionless utility maximization problem, i.e., Theorem 2.2 in \cite{KS99}, that $(\hvp^0,\hvp^1)\in\cA\big(x;\frac{\hY^1}{\hY^0}\big)$ and $\hY^0\in\cY\big(\widehat y(x);\frac{\hY^1}{\hY^0}\big)$ are the solutions to the frictionless primal and dual problem for $\frac{\hY^1}{\hY^0}$, if $\widehat y(x;\frac{\hY^1}{\hY^0})=\widehat y(x)$. To see the latter, we observe that $u(x)=v\big(\widehat y(x)\big)+x\widehat y(x)$ by Theorem \ref{mainthm} and therefore $v\big(\widehat y(x)\big)+x\widehat y(x)=u(x)\leq u\big(x;\frac{\hY^1}{\hY^0}\big)\leq v\big(\widehat y(x);\frac{\hY^1}{\hY^0}\big)+x\widehat y(x)$ by \eqref{int:rel2}. Since $v\big(\widehat y(x)\big)=E\big[V(\hY^0_T)\big]$, $E[\hvp^0_{T+1}\hY^0_T]=x\widehat y(x)$ and $\hY^0\in\cY\big(\widehat y(x);\frac{\hY^1}{\hY^0}\big)$, we obtain that $\widehat y(x;\frac{\hY^1}{\hY^0})=\widehat y(x)$, which completes the proof.
\ep 

\bp[Proof of Proposition \ref{lem:connection}]
As the solution $(\vp^0,\vp^1)\in\cA(x;\hS)$ to \eqref{p1} exists by the definition of a shadow price, we obtain that $\hS=(\hS)_{t=0}^T$ is arbitrage-free. 
To see this, suppose that there exists an arbitrage opportunity. Since we are in finite discrete time, the existence of an arbitrage opportunity is equivalent to the existence of an arbitrage opportunity $(\xi^0,\xi^1)$ in one period. Adding this (as in the proof of Theorem 3.3 in \cite{FS04})
to the solution $(\vp^0,\vp^1)\in\cA(x;\hS)$ to \eqref{p1} yields a strategy with a strictly higher expected utility, i.e., $E[U(\vp^0_{T+1})]<E\big[U\big((\vp^0+\xi^0)_{T+1}\big)\big]$. As $(\vp^0+\xi^0,\vp^1+\xi^1)\in\cA(x;\hS)$, this contradicts the assumption that $(\vp^0,\vp^1)$ is the solution to \eqref{p1} and hence proves that $\hS$ is arbitrage-free. By the fundamental theorem of asset pricing \cite{DMW90} this implies the existence of an equivalent martingale measure $Q$ for $\hS$ and therefore that $\cY(y;\hS)\ne\emptyset$ for all $y>0$. As $\hS$ is valued in the bid-ask spread $[(1-\lambda)S,S]$, any $(\vp^0,\vp^1)\in\cA(x)$ is also self-financing for $\hS$ without frictions ($\lambda=0$) and hence $\cA(x)\subseteq\cA(x;\hS)$. Since $\cA(x)\subseteq\cA(x;\hS)$, we obtain that $(Y^0,Y^1):=(Y,Y\hS)\in\cB(y)$ for all $Y\in\cY(y;\hS)$ and therefore similarly to \eqref{int:rel2}:
\be
v(y)=\inf_{(Y^0,Y^1)\in\cB(y)}E[V(Y^0_T)]\leq \inf_{Y\in\cY(y;\hS)}E[V(Y_T)]=: v(y;\hS).\label{int:rel3}
\ee
Moreover, as
$$u(x) = v\big(\widehat y(x)\big)+x\widehat y(x)\leq v\big(\widehat y(x;\hS)\big)+x\widehat y(x;\hS) \leq v\big(\widehat y(x;\hS);\hS\big)+x\widehat y(x;\hS) =u(x;\hS)=u(x),$$
it follows that $\widehat y(x)=\widehat y(x;\hS)$ and therefore that $(\widehat{Y}^0,\widehat{Y}^1):=(\hY,\hY\hS)\in\cB\big(\widehat y(x)\big)$ is the solution to the frictional dual problem \eqref{D1}, where $\hY\in\cY\big(\widehat y(x;\hS)\big)$ is the solution to its frictionless counterpart \eqref{d1} for $\widehat{S}$.
\ep 

\bp[Proof of Proposition \ref{lem:super}]
In analogy to the primal problem, we introduce a two-dimensional objective function 
$$\mathcal{V}(x,y):=\begin{cases} V(x)~&:\text{ $y\geq 0$ and $x>0$},\\
+\infty~&:\text{ else}
\end{cases}$$
for the dual problem \eqref{D1} that is then given by
\be
E[\mathcal{V}(Y^0_{T},Y^1_{T})]\to\max!,\qquad{(Y^0,Y^1)\in\cB(y)}.\label{D2}
\ee
For a dynamic formulation of \eqref{D2} we set $\cB(Y^0,Y^1;t)=\{(\widetilde{Y}^0,\widetilde{Y}^1)\in\cB(y)~|~(\widetilde{Y}^0_s,\widetilde{Y}^1_s)=(Y^0_s,Y^1_s)~\text{for $s=0,\ldots,t$}\}$ for $(Y^0,Y^1)\in\cB(y)$ and 
$$
\mathcal{V}_t(Y^0,Y^1):=\underset{(\widetilde{Y}^0,\widetilde{Y}^1)\in\cB(Y^0,Y^1;t)}{\essinf}E\big[\mathcal{V}(\widetilde{Y}^0_{T},\widetilde{Y}^1_{T})\big|\cF_t\big].\footnote{Note that the definition of $\cU_t$ and $\mathcal{V}_t$ is slightly asymmetric, as the set $\cA$ is -- in contrast to the set $\cB(y)$ -- stable under concatenation.}
$$
By the martingale optimality principle we then have $\cU_t(\hvp^0_t,\hvp^1_t)=E\big[\cU(\hvp^0_{T+1},\hvp^1_{T+1}\big)\big|\cF_t\big]$ for the solution $(\hvp^0,\hvp^1)$ to \eqref{P2} and $\mathcal{V}_t(\hY^0,\hY^1)=E\big[\mathcal{V}(\hY^0_{T},\hY^1_{T}\big)\big|\cF_t\big]$ for the solution $(\hY^0,\hY^1)$ to \eqref{D2}. Now, by Fenchel's inequality, we have
$$\cU\big(\psi^0+(\tvp^0_{T+1}-\tvp^0_{t}),\psi^1+(\tvp^1_{T+1}-\tvp^1_{t})\big)\leq \mathcal{V}(\hY^0_{T},\hY^1_{T})+\big(\psi^0+(\tvp^0_{T+1}-\tvp^0_{t})\big)\hY^0_{T}+\big(\psi^1+(\tvp^1_{T+1}-\tvp^1_{t})\big)\hY^1_{T}$$
for all $(\tvp^0,\tvp^1)\in\cA$ and $U(\hvp^0_{T+1})=\cU(\hvp^0_{T+1},\hvp^1_{T+1})=\mathcal{V}(\hY^0_{T},\hY^1_{T})+\hvp^0_{T+1}\hY^0_{T}+\hvp^1_{T+1}\hY^1_{T}$ for $(\hvp^0,\hvp^1)$ by part 3) of Theorem \ref{mainthm}. Taking conditional expectations and optimizing in the above equation gives
\be
\cU_t(\psi^0,\psi^1)\leq \mathcal{V}_t(\hY^0,\hY^1)+\psi^0\hY^0_{t}+\psi^1\hY^1_{t}\label{eq1}
\ee and
\be
\cU_t(\hvp^0_{t},\hvp^1_{t})= \mathcal{V}_t(\hY^0,\hY^1)+\hvp^0_{t}\hY^0_{t}+\hvp^1_{t}\hY^1_{t}\label{eq2}\ee
by the martingale optimality principle. In \eqref{eq1} we exploit that there either exists $(\vp^0,\vp^1)\in\cA(\widetilde{x})$ for some $\widetilde{x}>0$ such that $(\vp^0_{T+1},\vp^1_{T+1})=\big(\psi^0+(\tvp^0_{T+1}-\tvp^0_{t}),\psi^1+(\tvp^1_{T+1}-\tvp^1_{t})\big)$ and that $(\hY^0_{t}\vp^0_{t+1}+\hY^1_{t}\vp^1_{t+1})_{t=0}^T$ is a supermartingale with $E[\hY^0_{T}\vp^0_{T+1}+\hY^1_{T}\vp^1_{T+1}|\cF_t]\leq\hY^0_{t}\psi^0+\hY^1_{t}\psi^1$ or the inequality is trivially satisfied with $\cU_t(\psi^0,\psi^1)=-\infty$, if no such $(\vp^0,\vp^1)$ exists. Combining \eqref{eq1} and \eqref{eq2} then implies \eqref{eq:supdiff}, i.e., that
$$
\cU_t(\psi^0,\psi^1)\leq \cU_t(\hvp^0_{t},\hvp^1_{t})+(\psi^0-\hvp^0_t)\hY^0_{t}+(\psi^1-\hvp^1_{t})\hY^1_{t},
$$
a.s.~for all $(\psi^0,\psi^1)\in L^0(\Om,\cF_t,P;\R^2)$.
\ep

\bibliographystyle{abbrv}
\end{document}